\documentclass{iau} 
\usepackage{graphicx}

\title[Molecular Clouds in the CMZ] %% give here short title %%
{Temperature Evolution of Molecular Clouds in the Central Molecular Zone}

\author[Nico Krieger et al.]   %% give here short author list %%
{Nico Krieger$^{1}$, J\"urgen Ott$^2$, Fabian Walter$^{1,2}$, J.~M.~Diederik Kruijssen$^{3,1}$, Henrik Beuther$^1$ \and the SWAG team}

\affiliation{%$^\dagger$krieger mpia.de\\[\affilskip]
$^1$Max-Planck-Institute f\"ur Astronomie, Heidelberg, Germany\\[\affilskip]
$^2$National Radio Astronomy Observatory, Socorro, NM, USA\\[\affilskip]
$^3$Astronomisches Rechen-Institut, Universit\"at Heidelberg, Germany\\[\affilskip]
}

\pubyear{2016}
\volume{322}  %% insert here IAU Symposium No.
\jname{The Multi-Messenger Astrophysics of the Galactic Centre}
\editors{Roland Crocker, Steve Longmore, Geoff Bicknell, eds.}

\begin{document}

\maketitle

\begin{abstract}
We infer the absolute time dependence of kinematic gas temperature along a proposed orbit of  molecular clouds in the Central Molecular Zone (CMZ) of the Galactic Center (GC).
Ammonia gas temperature maps are one of the results of the ``Survey of Water and Ammonia in the Galactic Center'' (SWAG, PI: J. Ott); the dynamical model of molecular clouds in the CMZ was taken from Kruijssen et al. (2015).
We find that gas temperatures increase as a function of time in both regimes before and after the cloud passes pericenter on its orbit in the GC potential.
This is consistent with the recent proposal that pericenter passage triggers gravitational collapse.
Other investigated quantities (line width, column density, opacity) show no strong sign of time dependence but are likely dominated by cloud-to-cloud variations. 

\end{abstract}

The Central Molecular Zone (CMZ) in the Galactic Center (GC) contains large amounts of dense molecular gas, but observed star formation rates are lower than expected by a factor of $10-100$ (Longmore et al. 2013a).
Part of this gas is projected onto an $\infty$-like shape and has been modeled as open-ended streams wrapping around the GC (Longmore et al. 2013b; Kruijssen et al. 2015).
The streams repeatedly pass close (60 pc) to the GC where cloud collapse is expected to be triggered by tidal interactions.
Downstream of the near side pericenter, a sequence of advancing star formation (SF) tracers is present that may indicate a sequence of progressing star formation states (Longmore et al. 2013b).
We aim to understand the conditions of molecular gas, which is the fuel of SF, at the different stages of the sequence.

\begin{figure}[h!]
	\centering
	\includegraphics[width=\linewidth]{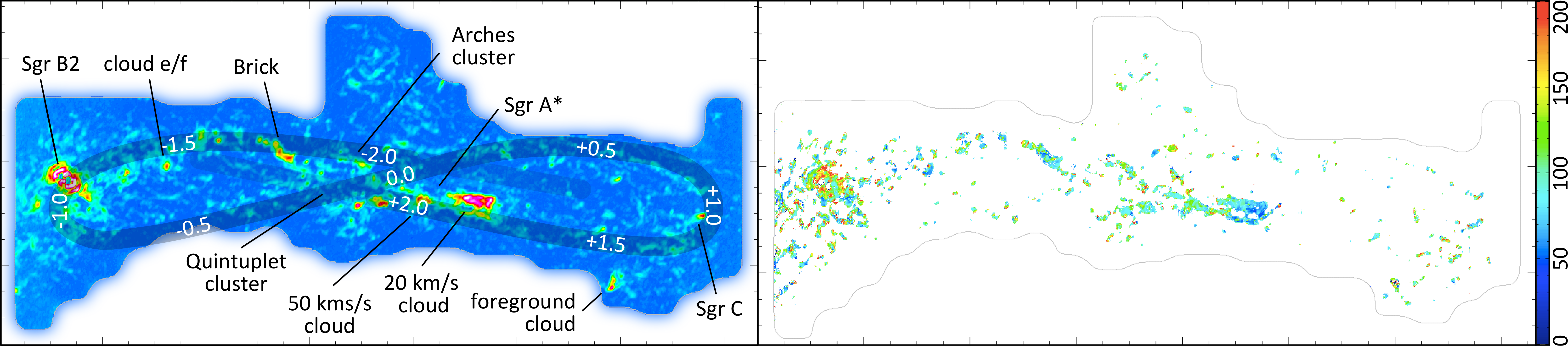}
	\caption{\emph{left:} Overview of the central CMZ. The NH$_3$~(1,1) peak flux map extends from -0.65$^\circ$ to +0.80$^\circ$ in Galactic longitude and $\pm 0.35^\circ$ in Galactic latitude. Gas streams are highlighted (dark band) and overplotted with time (white) since the far side pericenter passage in the model of Kruijssen et al. (2015). Pericenter passages occur at $\pm 2.0$\,Myr and 0.0\,Myr. The ``dust ridge'' spans between the ``Brick'' and Sgr B2. \emph{right:} NH$_3$~(2,2)-(4,4) kinetic temperature (T$_{24}$ in Kelvin) map of the CMZ derived from pixel-by-pixel fitting of the ammonia hyperfine structure (Krieger 2016a, 2016b). Blanked regions cannot be fitted due to low signal-to-noise ratios.}
	\label{CMZ and T24}
\end{figure}

\newpage
In the ``Survey of Water and Ammonia in the Galactic Center'' (SWAG, cf. overview by Ott et al. in this volume) targeted the CMZ at six meta-stable ammonia lines among a variety of other spectral lines.
Fig. \ref{CMZ and T24} (left panel) shows a peak intensity map of NH$_3$~(1,1) highlighting the ring-like gas streams and other relevant objects like molecular clouds and star clusters.
The right panel of Fig. \ref{CMZ and T24} represents the same view in kinetic gas temperature T$_{24}$ which can be derived from the relative line fluxes through hyperfine structure fits of NH$_3$~(2,2) and NH$_3$~(4,4) on a pixel by pixel basis (Krieger 2016a, 2016b).

\begin{figure}
	\centering
	\includegraphics[width=0.71\linewidth]{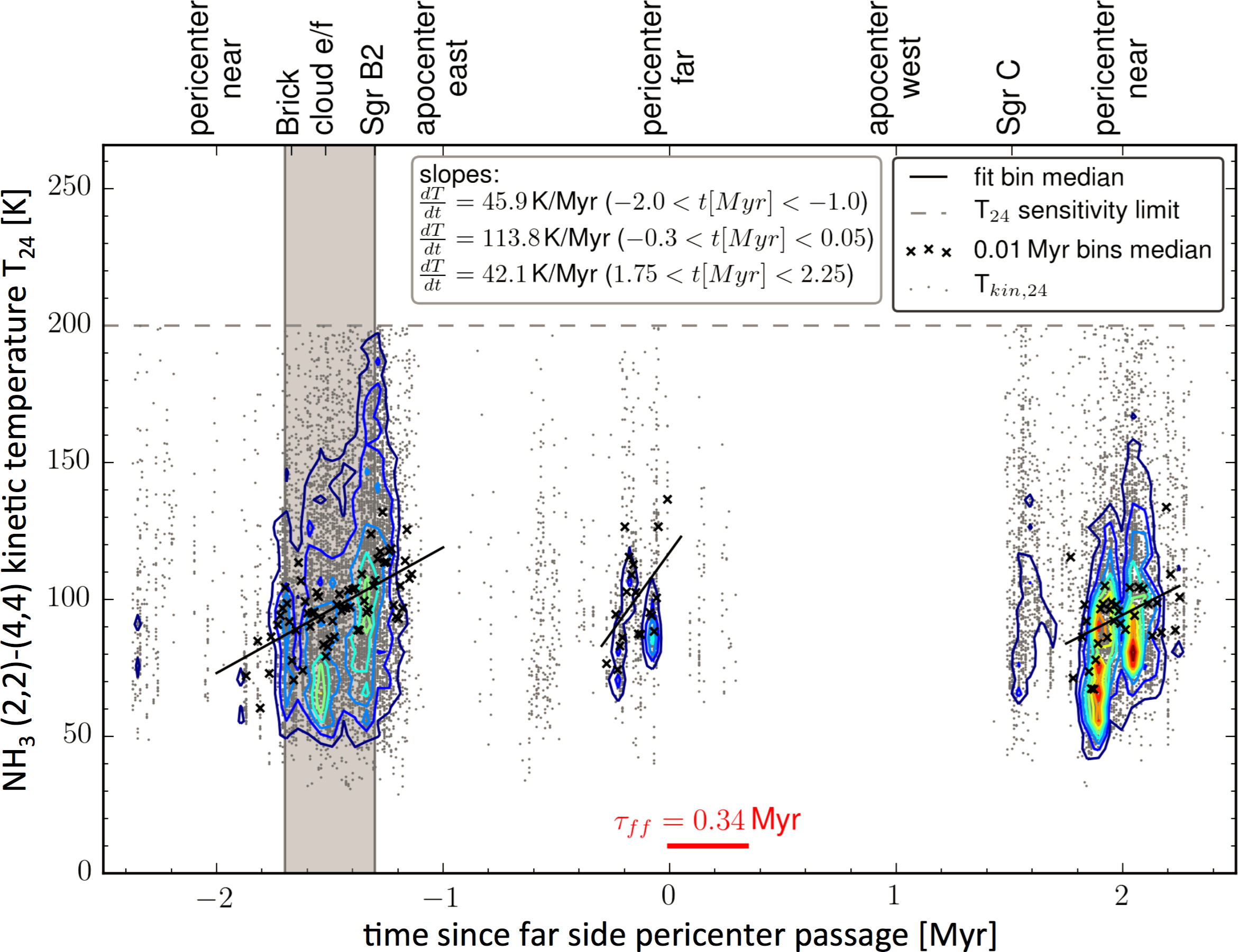}
	\caption{Kinetic ammonia temperature T$_{24}$ [NH$_3$ (2,2) to (4,4)] as function of time since far side pericenter passage. Individual temperature measurements are overplotted with data density contours, medians of bins of 0.01 Myr width (crosses) and linear fits to these medians.}
	\label{time vs T24}
\end{figure}

The kinematic model of Kruijssen et al. (2015, see marked orbit in fig. \ref{CMZ and T24}, left) allows us to assign time stamps to CMZ clouds if their 3D position (longitude, latitude, line-of-sight velocity) is known.
Thus, mapped quantities like temperature can be plotted as a function of time as in Fig.~\ref{time vs T24} for kinetic gas temperature T$_{24}$.
Temperatures increase in the dust ridge (-1.7 to -1.3 Myr) and near side pericenter passage (+1.8 to +2.0 Myr) as shown by the fits at consistent slopes of $\sim 46$ K/Myr and $\sim 42$ K/Myr.
At 0.0 Myr (far side pericenter), we find a similar trend, although the statistical basis is weak because of the short sampled time range and few measurements due to low SNR.

We also find quantitatively similar results for kinetic ammonia temperatures T$_{45}$ [NH$_3$~(4,4) to (5,5)] and T$_{36}$ [NH$_3$~(3,3) to (6,6)]. 
The physical reasons for differences between temperature slopes of different temperature tracers are currently unknown but likely reflect different conditions in the clouds.

Other mapped quantities (column density, opacity, line width) are dominated by variations of cloud structure and do not show consistent evolution with time.

\end{document}